\input{aipcheck}

\documentclass[
    ,final            
  ]
  {aipproc}

\layoutstyle{6x9}

\begin{document}

\title{Dark-energy equation of state: \\ how far can we go from $\Lambda$?}


\author{Hrvoje \v{S}tefan\v{c}i\'{c}}{
  address={Departament d'Estructura i Constituents de la Mat\`{e}ria, Universitat de Barcelona, \\ Av. Diagonal 647,  08028 Barcelona, Catalonia, Spain},
  altaddress={Theoretical Physics Division,
Rudjer Bo\v{s}kovi\'{c} Institute,
   P.O.Box 180, HR-10002 Zagreb, Croatia \footnote{permanent address}}
}



\begin{abstract}
 The equation of state of dark energy is investigated to determine how much it may deviate from the equation of state of the cosmological constant (CC). Two aspects of the problem are studied: the "expansion" around the vacuum equation of state and the problem of the crossing of the cosmological constant boundary.
\end{abstract}

\maketitle


\section{Introduction}

\label{intro}

A number of mechanisms have been proposed to explain the present accelerated expansion of the universe, which is now well established by diverse cosmological observations such as supernovae of the type Ia \cite{SNIa}, the large-scale structure of the universe \cite{LSS}, or anisotropies of the microwave background radiation \cite{CMB}. The most notable proposal towards the explanation of the accelerated expansion of the universe is the existence of the cosmic component characterized by the negative pressure, the so-called {\em dark energy} (DE) \footnote{Interesting alternatives to dark energy comprise braneworld models and modifications of gravity at cosmological scales.} \cite{DE}. It is customary to describe dark energy as a cosmic fluid defined by its equation of state (EOS) of the form $p_d=w \rho_d$, where $\rho_d$ and $p_d$ stand for dark-energy density and pressure, respectively. A key goal in the analyses of the present and the future observational data is establishing whether $w$ is constant or whether it evolves with the expansion of the universe. The analyses of the present observational data, assuming a constant $w$, restrict $w$ to a relatively narrow interval around $w=-1$, the value characterizing the cosmological constant $\Lambda$ \cite{CMB}. Other analyses of the observational data, allowing for the variability of $w$ with the redshift, do not impose so stringent bounds and also allow for (or even mildly favor) the phenomenon of the cosmological constant boundary crossing (i.e., crossing of the $w=-1$ line) at a small redshift \cite{crossobs}. The results of both of these analyses provide strong motivation for the study of dark-energy models which are, at least in some redshift interval, characterized by the EOS close to the CC EOS. We pursue this study in two directions: the first one is the study of corrections to the CC EOS (a sort of "expansion" around the vacuum EOS), whereas the second one comprises the study of the CC boundary crossing in terms of the {\em implicit} DE EOS.
In the second section we consider a class of models which represent an "expansion" around the vacuum EOS. In the third section we discuss the form of the dark-energy EOS necessary for the dark-energy model to exhibit the crossing of the CC boundary. We conclude in the last section.


\section{"Expansion" around the vacuum equation of state for dark energy}

\label{sect1}

Since the $\Lambda$CDM model gives a phenomenologically simple description of the available observational data and it has so far passed all observational tests, it is reasonable to expect that the dynamical features of dark energy, if they exist, should not be too prominent and the dynamics of dark energy, at least in the redshift range amenable to SNIa observations, should be close to the cosmological constant. In such a situation, it is useful and convenient to study the dark-energy EOS as an "expansion" around the CC EOS in the form $p_{d}=-\rho_{d}-f(\rho_d)$ \cite{odin1} (or equivalently as adding a correction to the CC EOS), where $p_d$ and $\rho_d$ are dark-energy pressure and density, respectively. A particularly interesting model is obtained when $f$ is a powerlike function of $\rho_d$, i.e., for the dark-energy EOS of the form \cite{odin1,PRDmoj}
\begin{equation}
\label{eq:oureos}
p_{d} = -\rho_{d} - A \rho_{d}^{\alpha} \, .
\end{equation}
This dark-energy model is conceptually simple, it can describe both quintessencelike and phantomlike forms of dark energy, it is highly analytically tractable and provides a wealth of physical phenomena in different parameter regimes \cite{PRDmoj,odin2}. The equation of continuity for DE
\begin{equation}
\label{eq:scalgen}
d \rho_{d} + 3(\rho_{d}+p_{d}) \frac{d \, a}{a} = 0 \, ,
\end{equation}
leads to the law of scaling with the scale factor for the model (\ref{eq:oureos}),
\begin{equation}
\label{eq:scalour}
\rho_{d}=\rho_{d,0} \left( 1 + 3 \tilde{A} (1-\alpha) \ln \frac{a}{a_{0}} \right)^{1/(1-\alpha)} \, ,
\end{equation}
where we have introduced an abbreviation $\tilde{A} = A \rho_{d,0}^{\alpha-1}$.
%
In a cosmological model which contains a nonrelativistic matter component $\rho_m$ except the dark-energy component (\ref{eq:oureos})  the dynamics of the scale factor $a$ is determined by the Friedmann equation
\begin{equation}
\label{eq:Fried}
\left( \frac{\dot{a}}{a} \right)^2 + \frac{k}{a^2} = \frac{8 \pi G}{3} (\rho_{d}+\rho_{m}) \, .
\end{equation}
In the regime where the DE density is much larger than the curvature term $k/a^2$ and $\rho_m$ it is possible to obtain an analytic expression for the time evolution of the scale factor. For $\alpha \neq 1/2$, the expression is
\begin{equation}
\label{eq:aodt}
\left( 1 + 3 \tilde{A} (1-\alpha) \ln \frac{a_{1}}{a_{0}} \right)^{\frac{1-2\alpha}{2(1-\alpha)}}
- \left( 1 + 3 \tilde{A} (1-\alpha) \ln \frac{a_{2}}{a_{0}} \right)^{\frac{1-2\alpha}{2(1-\alpha)}}
=\frac{3}{2} \tilde{A} (1-2\alpha) \Omega_{d,0}^{1/2} H_{0} (t_{1}-t_{2}) \, .
\end{equation}
Here  $\Omega_{d,0}=\rho_{d,0}/\rho_{c,0}$, where $\rho_{c,0} = 3 H_{0}^2/(8 \pi G)$ is the critical energy density
at the present epoch.
For $\alpha = 1/2$, the evolution of $a$ becomes
\begin{equation}
\label{eq:aodtspec}
\ln \frac{1 + \frac{3}{2} \tilde{A} \ln \frac{a_{1}}{a_{0}}}
{1 + \frac{3}{2} \tilde{A} \ln \frac{a_{2}}{a_{0}}} = \frac{3}{2} \tilde{A} \Omega_{d,0}^{1/2} H_{0} (t_{1}-t_{2}) \, .
\end{equation}

\begin{figure}[floatfix]
\resizebox{0.45\textwidth}{!}{\includegraphics{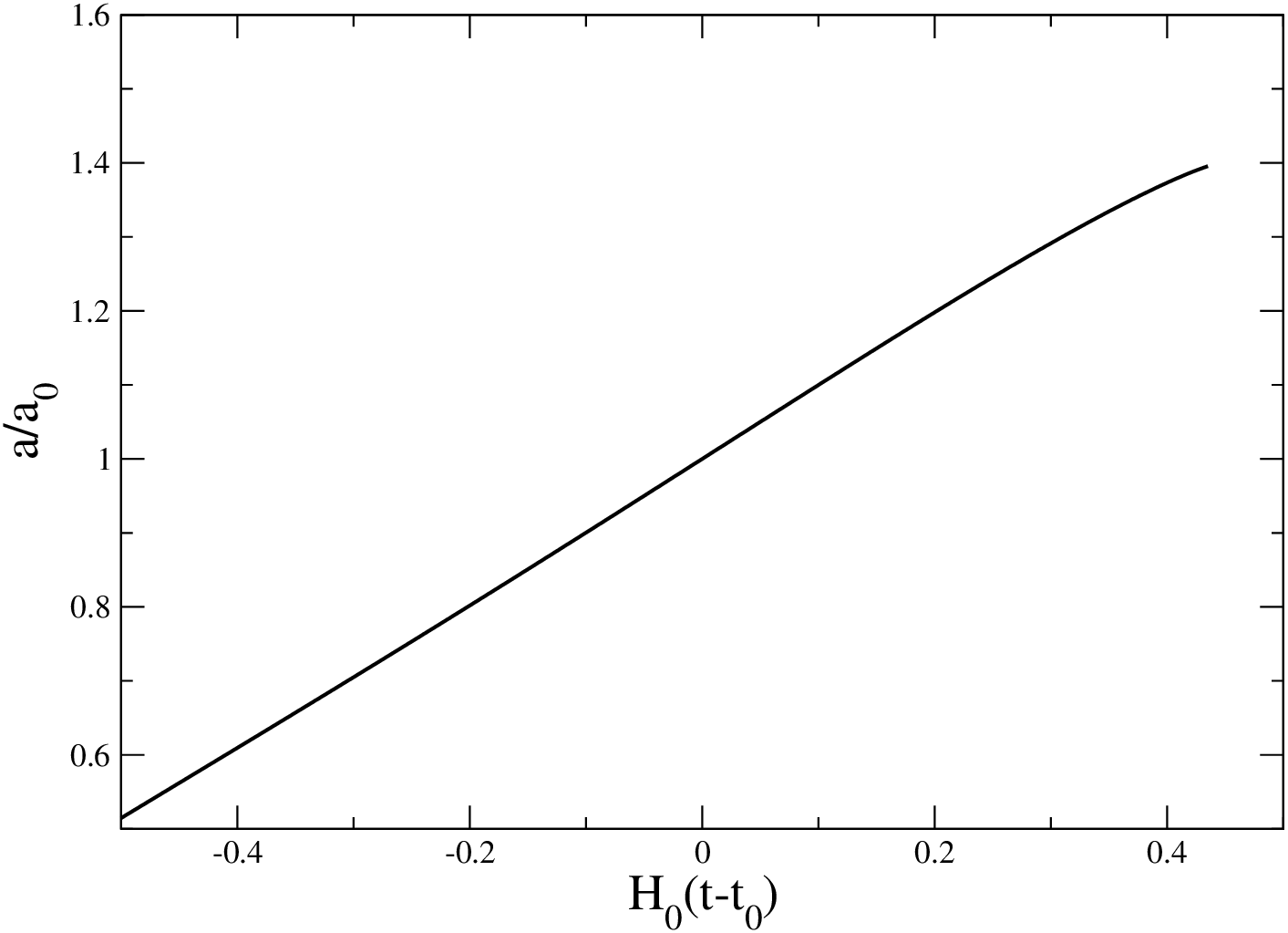}}
\hspace{5mm}
\resizebox{0.45\textwidth}{!}{\includegraphics{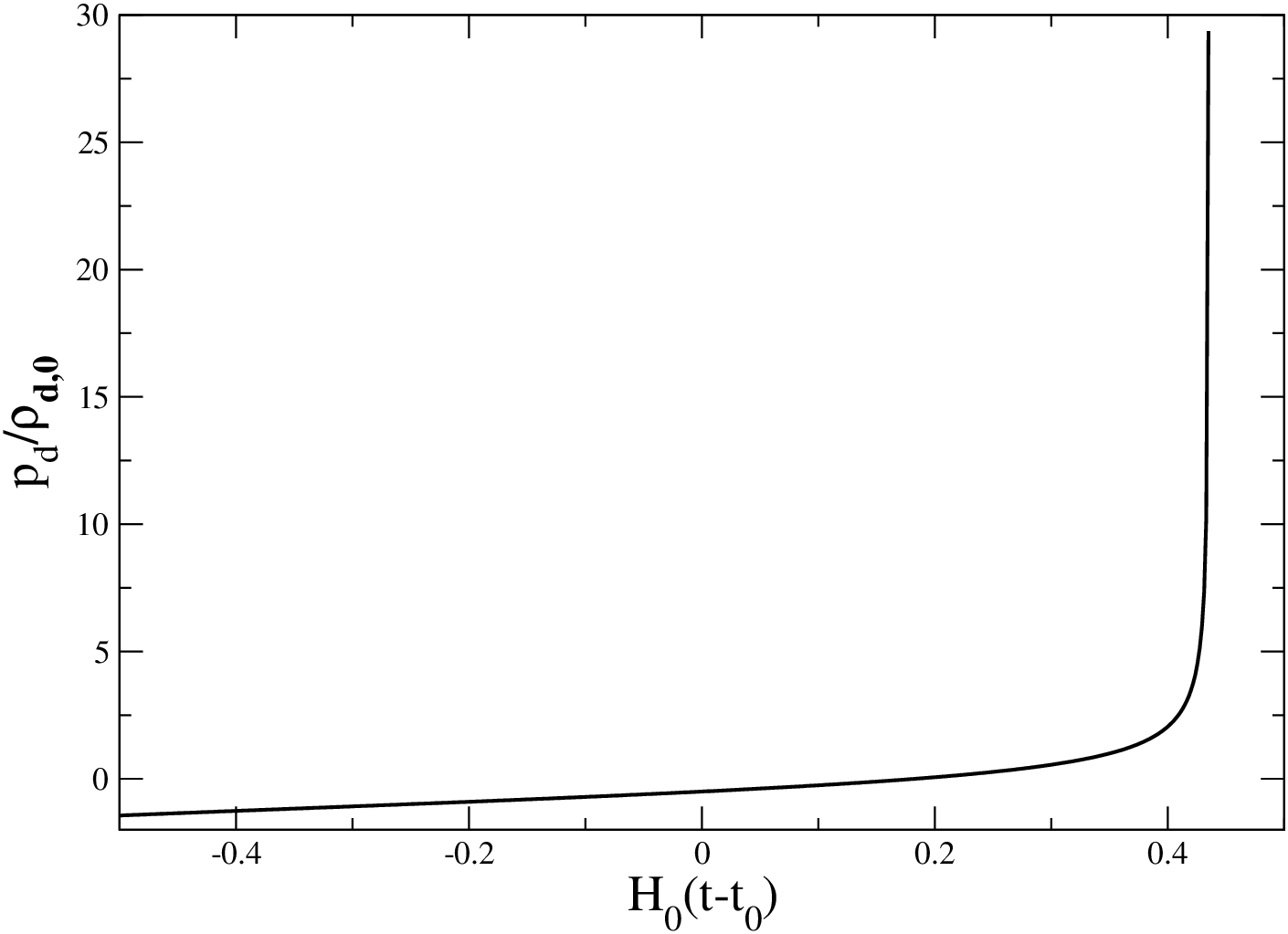}}
\caption{\label{fig:pressureBar} The dependence of the scale factor $a$ (left plot) and the dark energy pressure (right plot) on cosmic time for $\Omega_{d,0}=0.7$, $\Omega_{m,0}=0.3$, $\tilde{A}=-0.5$ and $\alpha=-1$. The divergence of the dark energy pressure at finite time and scale factor is clearly demonstrated.   }
\end{figure}

The analysis of the dependence of cosmological dynamics on the parameters $(\tilde{A}, \alpha)$ reveals a diversity of very interesting phenomena \cite{PRDmoj,AlcStef}, especially related to the asymptotic expansion and the fate of the universe. For $\tilde{A} > 0$ and $\alpha > 1$, the universe has a singular ending at finite time and at {\em a finite value of the scale factor}. For  $\tilde{A} > 0$ and $1/2 < \alpha < 1$, the universe encounters the singularity of the ``big rip" type at finite time. For $\tilde{A} > 0$ and $\alpha < 1/2$, the asymptotic expansion of the universe is nonsingular. For $\tilde{A} < 0$, the most interesting phenomenon occurs for $\alpha < 0$. In this parametric regime the universe encounters a {\em sudden future singularity} introduced by Barrow \cite{Barrow}. Namely, at the singularity, which happens at finite time, the scale factor, dark-energy and total-energy density and the Hubble parameter are finite, whereas the dark energy pressure and the acceleration diverge $p_d \rightarrow \infty, \ddot{a} \rightarrow - \infty$, as depicted in Fig. \ref{fig:pressureBar}. Furthermore, for $\tilde{A} < 0$ and $\alpha < 1$, the acceleration in the model is transient \cite{AlcStef}. Comparing the dynamics of the model against the observational data in principle provides an exciting possibility that one of the scenarios described above might be selected by the data. Such a statistical analysis \cite{AlcStef}, however, reveals that the parameter $\tilde{A}$ can be somewhat constrained ($-0.06 < \tilde{A} < 0.12$ at 95.4\% c.l.), whereas the present data provide no reliable constraint on $\alpha$. Hopefully, the forthcoming observational data will provide more stringent parameter constraints and scenario selection.

\section{Cosmological constant boundary crossing: a description in terms of the implicit dark-energy equation of state}

\label{sect2}

In the observational testing of the possible dynamical nature of dark energy, the DE parametrizations which allow for the redshift dependence of $w$ are of special interest. The analyses using this kind of parametrizations \cite{crossobs} allow for or even mildly favor the phenomenon of the CC boundary crossing. Namely, $w(z)$ acquires the value -1 at some finite $z$. While the modeling of this phenomenon \cite{cross1} usually employed multiple scalar fields or described it as an effective phenomenon in a more complex model \cite{cross2}, here we discuss the phenomenon from the viewpoint of the DE EOS for a single dark-energy component which is separately conserved, i.e., noninteracting with other components \cite{PRDcross}. We are interested in what form the DE EOS must have for dark energy to exhibit the CC boundary crossing.

We start with an example of the DE model which is explicitly constructed to exhibit the crossing. The scaling of the DE density with the scale factor is
\begin{equation}
\label{eq:denmod1}
\rho_d = C_{1} \left( \frac{a}{a_{0}} \right)^{-3(1+\gamma)} +
C_{2} \left( \frac{a}{a_{0}} \right)^{-3(1+\eta)} \, ,
\end{equation}
where in general $\gamma > -1$ and $\eta < -1$.
%
%
%
The EOS of the model defined by (\ref{eq:denmod1}) is obtained in a straightforward manner
\begin{equation}
\label{eq:eosdetmod1}
\frac{p_d -\eta \rho_d}{(\gamma-\eta) C_{1}} =
\left( \frac{\gamma \rho_d - p_d}{(\gamma-\eta) C_{2}} \right)^{(1+\gamma)/(1+\eta)}
\, .
\end{equation}
The main feature of this EOS is that it is {\em implicitly} defined. This explicitly constructed example suggests that the DE EOS which can exhibit the CC boundary crossing should be defined implicitly. A more detailed understanding of the crossing mechanism, however, requires a study of a broader class of models. Let us consider a generalized model \cite{PRDcross} with an EOS
\begin{equation}
\label{eq:eosgenmod1}
A \rho_d + B p_d = (C \rho_d + D p_d)^{\alpha} \, .
\end{equation}
It is important to notice that (\ref{eq:eosdetmod1}) is a special case of (\ref{eq:eosgenmod1}), so the generalized model is capable of exhibiting the crossing at least for some parameter values.
Starting from (\ref{eq:eosgenmod1}), the conservation of the DE component yields the evolution of $w$ with the scale factor
\begin{equation}
\label{eq:evolw1}
\left( \frac{\alpha}{(F+w)(1+w)}-\frac{1}{(E+w)(1+w)} \right) dw = 3(\alpha - 1)
\frac{da}{a} \, ,
\end{equation}
where $E=A/B$ and $F=C/D$. Its solution for the most interesting case $E \neq -1$ and $F \neq -1$, where $E$ and $F$ lie on the opposite sides of the CC boundary, has the following form
\begin{equation}
\label{eq:solw}
\left| \frac{w+F}{w_{0}+F} \right|^{\alpha/(1-F)}
\left| \frac{w+E}{w_{0}+E} \right|^{-1/(1-E)}
\left| \frac{1+w}{1+w_{0}} \right|^{1/(1-E)-\alpha/(1-F)}
= \left( \frac{a}{a_{0}} \right)^{3(\alpha-1)} \, .
\end{equation}
%
%
%
%
Equivalently, the equation for the evolution of $w$ can be written as
\begin{equation}
\label{eq:wstar}
\frac{w+\frac{\alpha E - F}{\alpha - 1}}{(F+w)(E+w)(1+w)} dw = 3 \frac{da}{a} \, .
\end{equation}
Both (\ref{eq:solw}) and (\ref{eq:wstar}) clearly demonstrate that whenever there exists the term corresponding to the CC boundary ($w=-1$) in these equations, the boundary cannot be crossed at finite value of $a$. This term must be removed from (\ref{eq:solw}) and (\ref{eq:wstar}) for the crossing to happen. Mathematically, this can be achieved for $\frac{\alpha E - F}{\alpha - 1} = 1$, i.e., formally, the mecahnism of the crossing is the cancellation of the term corresponding to the CC boundary.

\begin{figure}
\centerline{\resizebox{0.8\textwidth}{!}{\includegraphics{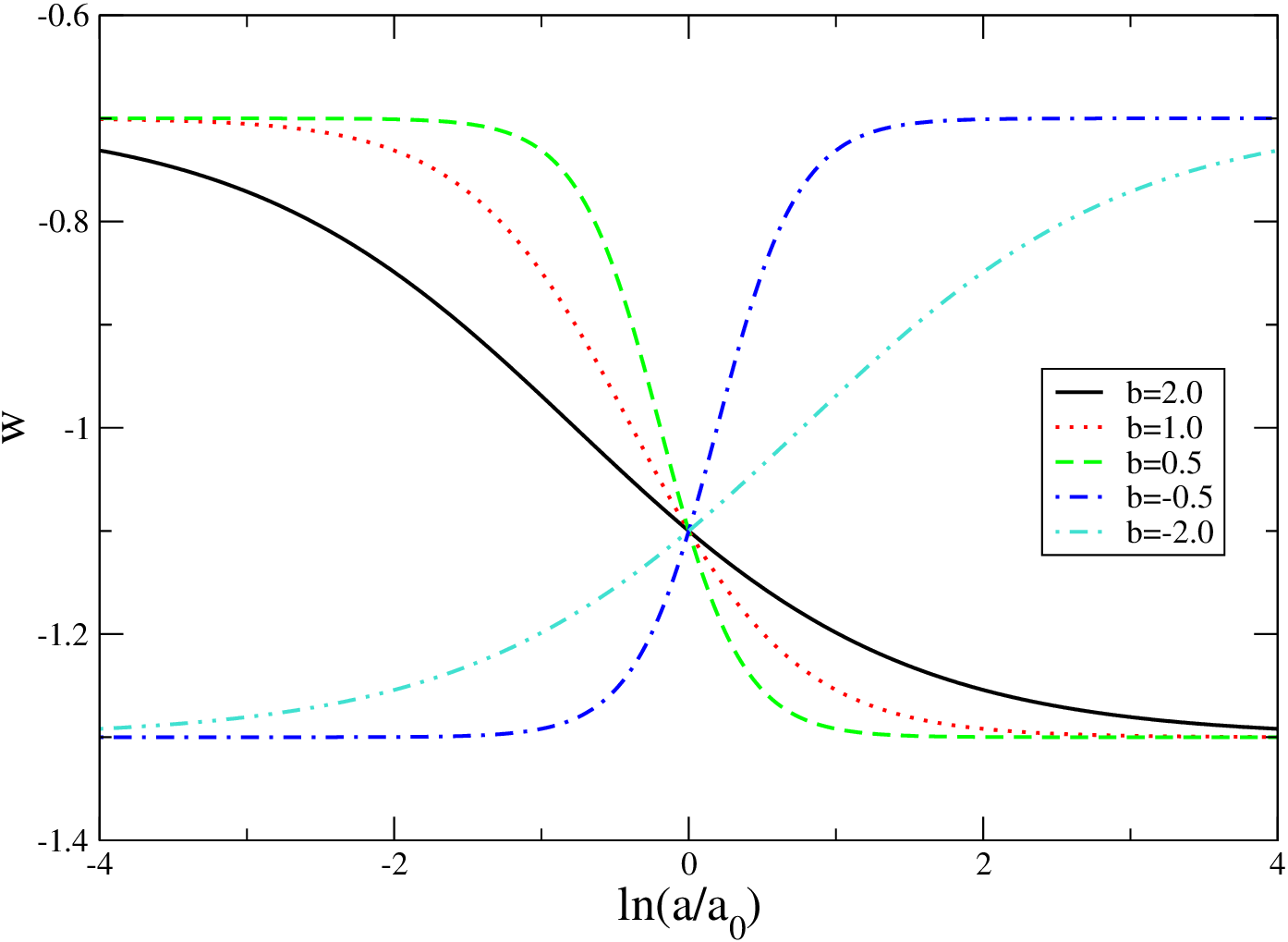}}}
\caption{\label{fig:mod2} The dependence of the EOS parameter $w$ of the model (\ref{eq:densitymod2}) on the scale factor $a$ for different values of the parameter $b$ and $w_0=-1.1$, $\gamma=-0.7$, and $\eta=-1.3$ .
}
\end{figure}

A more general DE model exhibiting the CC boundary crossing can be constructed. The scaling of its energy density is
\begin{equation}
\label{eq:densitymod2}
\rho_d = \left( C_{1} \left( \frac{a}{a_{0}} \right)^{-3(1+\gamma)/b} +
C_{2} \left( \frac{a}{a_{0}} \right)^{-3(1+\eta)/b} \right)^{b} \, ,
\end{equation}
and the scaling of the corresponding parameter of its EOS is depicted in Fig. \ref{fig:mod2}. It is also characterized by the {\em implicitly} defined EOS
%
%
%
%
\begin{equation}
\label{eq:eosdetmod2}
\frac{p_d -\eta \rho_d}{(\gamma-\eta) C_{1}} =
\rho^{((1-b)(\gamma - \eta))/(b(1+\eta))}
\left( \frac{\gamma \rho_d - p_d}{(\gamma-\eta) C_{2}} \right)^{(1+\gamma)/(1+\eta)}
\, ,
\end{equation}
and the consideration of the corresponding generalized DE model with the EOS of the type
\begin{equation}
\label{eq:eosmodel2}
A \rho_d + B p_d = (C \rho_d + D p_d)^{\alpha} (M \rho_d + N p_d)^{\beta} \, ,
\end{equation}
reveals that the same cancellation mechanism as in (\ref{eq:eosgenmod1}) is responsible for the crossing \cite{PRDcross}.
%
%
%
Moreover, using the insight provided by the preceding examples, it is possible to show \cite{PRDcross} that even the complex DE EOS of the form
\begin{equation}
\label{eq:ton}
A \rho_d^{2n+1} + B p_d^{2n+1} = (C \rho_d^{2n+1} + D p_d^{2n+1})^{\alpha} \, ,
\end{equation}
for $n \ge 0$ may describe the CC boundary crossing.
%
%
%

After the analysis of the particular classes of DE models exhibiting the CC boundary crossing, one may attempt to define more general conditions on the DE models for the description of the crossing. For a very general class of DE models which may be parametrically defined in terms of $w$
\begin{equation}
p_d=p_d(w) \, , \;\;\;\; \rho_d=\rho_d(w) \, ,
\end{equation}
%
it is possible to show that if the function $g(w)$ defined as
\begin{equation}
\label{eq:defg}
\frac{1}{\rho_d} \frac{d\rho_d}{d w} = (1+w) g(w) \,
\end{equation}
has the following property:
\begin{equation}
\label{eq:gprop}
\lim_{w \rightarrow -1} g(w) = \mathrm{ finite (nonzero)} \, ,
\end{equation}
then the CC boundary crossing can happen at a finite scale factor and the dynamics of $w$ is determined by the following general equation:
\begin{equation}
g(w) d w = -3 \frac{d a}{a} \, .
\end{equation}
All functions $g(w)$ satisfying (\ref{eq:defg}) and (\ref{eq:gprop}) lead to DE models which at some $a$ cross the CC boundary. The comparison against the observational data determines which of these mathematical possibilities are acceptable as the description of our universe.

\section{Conclusions}

The presented models of dark energy are used as tools to study an important question for the dynamical dark-energy models in general: how close must their EOS be to the CC one? The dynamics of the models studied in the second section shows that even if the DE EOS is presently close to the vacuum CC, the fate of the universe may be quite different from the deSitter asymptotic expansion of the $\Lambda$CDM models. Alternatively, the DE models with the implicitly defined EOS may describe the CC boundary crossing where the DE EOS also stays close to the CC EOS in a certain redshift interval, but its asymptotic behavior may also be very different from the behavior of the $\Lambda$CDM model. Therefore, there are many DE models presently close to the CC with a very different future dynamics which will, hopefully, be further constrained by the forthcoming observational data.

\label{concl}


\begin{theacknowledgments}
The author acknowledges the support of the Secretar\'{\i}a de Estado de Universidades e Investigaci\'{o}n of the Ministerio de Educaci\'{o}n y Ciencia of Spain within the program ``Ayudas para movilidad de Profesores de Universidad e Investigadores espa\~{n}oles y extranjeros". This work has been supported in part by the MEC and the FEDER under project 2004-04582-C02-01 and by the Dep. de Recerca de la Generalitat de Catalunya under contract CIRIT GC 2001SGR-00065. The author would like to thank the Departament E.C.M. of the Universitat de Barcelona for hospitality. The author would like to thank J. Alcaniz for a fruitful collaboration.
The author is also partially supported by the Ministry of Science, Education and Sport of the Republic of Croatia.
\end{theacknowledgments}



\end{document}